# The Model of the Mechanism of the VLF/LF, ULF Emission Influence on the Local Meteorological Processes during the Earthquake Preparation Period

M.K. Kachakhidze, Z.A. Kereselidze, N.K. Kachakhidze

St. Andrew The First-Called Georgian University of The Patriarchy of Georgia, Tbilisi, Georgia

*Correspondence to:* M. K. Kachakhidze    manana_k@hotmail.com

**Abstract**

The satellite and ground-based observations proved that VLF/LF and ULF electromagnetic emissions take place during earthquake preparation period in the focal area of an earthquake.

This work, using a model (analogous circuit model) of self- generated electromagnetic oscillations of separate segments of the LAI system, considers the influence of VLF/LF and ULF emission in the pre-earthquake period on a local sector of the global atmospheric electrical circuit. In order to distinguish effects causing the local particularities of the meteorological condition in the focal area of the incoming earthquake, the work describes the possibility of atmospheric electric field inversion and oscillation effects influence on the processes taking place in the cloud. It shows possibility of generation of the atmospheric electromagnetic radiation in the $10-10^6$ Hz frequency diapason in the conditions corresponding to "bad weather". In order to create a model of atmospheric effect of telluric electric current in the incoming earthquake focus we used classical analytical expressions of electrodynamics and considered a generalized model of geoelectric anomaly.

## 1. Introduction

In order to show possible mechanism of the ground-based electromagnetic VLF/LF and ULF emission generated in the focal area of the incoming earthquake during accumulation of tectonic stress we suggested a model of electromagnetic self- generated oscillations of the lithosphere-atmosphere-ionosphere (LAI) system (analogous circuit model) (Kachakhidze et al., 2011).

The present work shows that the frequency spectrum of Earth VLF electromagnetic emission is analytically determined by a linear parameter characterizing the upcoming earthquake. The value of the parameter changes from the length characterizing the preliminary cracks to the final main fault length:

$$\omega = \beta \frac{c}{l} \qquad (1)$$

where $\omega$ -is the main frequency, $l$ - is the main fault length in the earthquake focus, $c$ -is the light velocity, and $\beta$ - is the characteristic coefficient of geological medium, which approximately equals to 1. However, it should be determined individually for each seismically active region or for a local segment of the lithosphere.

As a result of retro-analysis of Earth electromagnetic emission spectrum in the case of emission spectrum monitoring in the period that precedes the earthquake it is possible to determine, with certain accuracy, the time, location and magnitude of an upcoming earthquake simultaneously.

There are a lot of records stating that before or after an earthquake in different regions different geophysical phenomena take place (Changing of intensity of electro-telluric current in the focal area; perturbations of geomagnetic field in forms of irregular pulsations or regular short-period pulsations; VLF electromagnetic emission is observed on the Earth's surface; perturbations of atmospheric electric field; cloud anomalies; irregular changing of characteristic parameters of the lower ionosphere: plasma frequency, electron concentration, height of D-layer etc.). For instance it is very interesting: irregular perturbations reach the upper ionosphere, namely F2-layer, for 2-3 days before the earthquake; intensity of electromagnetic emission increases in the upper ionosphere in several hours or tenths of minutes before earthquake and sometimes it causes lighting. However, any effort to connect them with the process of tectonic stress accumulation has not succeeded yet. These phenomena are especially frequent for $M \geq 5.0$ magnitude earthquakes (Uyeda et al., 2000, Kachakhidze et al., 2009, Genzano et al., 2009, Takeuchi et al., 2012, Zhang et al., 2013, Li et al., 2013, Guangmeng et al., 2013, Hayakawa et al., 2013, Boudjada et al., 2013) It is logical to suppose that the initiator of these phenomena must be perturbations of geophysical fields caused by the earthquake preparation process. Therefore, revealing and modeling of action of mechanisms causing these perturbations is a quite actual task.

In the focal area of the incoming earthquake some geophysical phenomena are observed, generation of which, at a glance, is not connected with the increase of the tectonic stress. In this regard local meteorological phenomena in the epicenter of the incoming earthquake are especially interesting. In order to analyze these phenomena we will use analogous circuit model (Kachakhidze et al., 2011) below.

## 1.1. Influence of the earthquake preparing process on the local meteorological condition

There are numerous ground – based and satellite data stating that before an earthquake, in the Earth atmosphere in parallel with generation of VLF/LF and ULF emission, in the increased tectonic stress zone, on the level of the lower ionosphere the so-called TEC (Total Electron Content) anomalies take place and local meteorological effects develop (De Agostino et al., 2011, Guangmeng et al., 2013).

Due to ambiguity of the phenomena connected with this problem, it is natural that there are different view points on the mechanism forming the cause-effect relationship between the Earth and atmosphere phenomena. Especially actual is the problem of separating the Earth electromagnetic emission from the magnetospheric one. This problem makes impossible to reliably determine cause-effect relationship between the seismic and atmospheric (ionospheric) phenomena (Masci, 2013). However, in the weakly ionized lower ionosphere medium with the influence of the VLF electromagnetic emission the mechanism of electrons concentration variation is generally obvious. But the role of this emission in meteorological phenomena is quite vague. It is neither qualitatively explained nor strengthened by quantitative data. It is logical that, in the first place, it is caused by variability of the electromagnetic image corresponding to the incoming earthquake focus caused by influence of various Earth and space factors. The electromagnetic effects caused by these factors are often overlapped by one another during the process of earthquake preparation. For example, it has been many times recorded that in the area of an incoming earthquake epicenter some TEC anomalies are observed in the ionosphere, whereas in the Earth some intensity variation of electro-telluric currents take place. It is often followed by different types of regular and irregular geomagnetic pulsations. However, these phenomena are not yet considered as reliable electromagnetic precursors for an earthquake (Masci, 2013). Like VLF electromagnetic emission in the lower ionosphere in the area of incoming earthquake epicenter the perturbations of the atmospheric electric field potential gradient and the parameters of the lower ionosphere (D-layer height, electron concentration, plasmic frequency) are regularly observed. The response of these perturbations sometimes reaches upper layers of the ionosphere.

Moreover, there are some data on local weather variation during an earthquake preparation period expressed in increased cloudiness (Guangmeng et al., 2013). Taking this into account a supposition was made that there is a link between local meteorological phenomena and Earth electromagnetic effects (Takeuchi et al. 2012) However, unambiguous prove for this supposition is prevented by permanent activity of the sun radiation as a global source of weather formation. Besides taking this factor into consideration, other local factors acting stationary on the meteorological condition, for example, the influence of the Earth surface orography and water reservoirs must be evaluated. These factors, apart from the Earth VLF electromagnetic emission may take part in formation of local weather. Namely, it is known that local factors direct influence on atmospheric thermodynamic conditions as well as on the electric fields.

Like local geographic-hydrologic conditions we should not exclude possibility of activation of one more source perturbing the atmospheric electric field: it is known that intensification of telluric electric currents may be caused by the influence of space as well as Earth sources. It is natural, that probability of significant electro-telluric effect must increase in the incoming earthquake focus due to the processes taking place in the area, e.g. due to increase of deep electric polarization effect. Thus, meteorological effect of the telluric current, in the case of its existence, must be especially distinguished in the earthquake preparation process. Consequently, for qualitative analysis of this phenomenon, an analogical circuit model, highlight of which is the local inversion of the atmospheric electric field, may appear useful (Kachakhidze et al., 2011). Therefore, presenting of deep polarization mechanism obviously at its most is important for telluric current as for increasing of reliability of one of the possible electromagnetic indicators for the earthquake preparation process.

## 2. Discussion

### 2.1. Changings of intensity of the electro-telluric current in an earthquake preparation area

The Earth is a condenced medium with specific properties where different thermodynamic and electromagnetic effects may develop. It is known that some certain time earlier before an earthquake in its focal area a process of cracking of the Earth's crust and unification of the cracks into a main fault take place, which gradually becomes more intense. Seemingly, the process of formation of the main fault (unification of cracks) is expressed in changings of Earth ULF and VLF electromagnetic emission spectrum (Kachakhidze et al., 2013).

It is natural that cracking process, together with electromagnetic emission, may be accompanied by acoustic and acoustic-gravitational wave generation (Takeuchi et al., 2012).

In parallel with formation of micro-cracks the electric dipoles will appear on their surface, generation of which takes place in an area of telluric currents. Therefore, according to (Kachakhidze et al., 2011) in the incoming earthquake focus a thermoionized channel may form, by influence of which positive charges are induced on the Earth's surface. According to this model, in case there is no telluric current in a medium, then, as soon as the thermoionized channel is formed a local electric field will form, which must be directed from the thermoionized channel (the surface of the unified cracks) to the Earth's surface. Therefore, the segment of the Earth containing the incoming earthquake focus, where tectonic stress in earthquake preparation period anomalously increases, must have positive electric potential towards atmosphere. It is noteworthy that the presence of such effect in nature is proved in the papers (Bleier et al., 2009; Eftaxias et al., 2009).

Due to the telluric current, by the influence of the thermoionized channel, the above mentioned current will become perturbed, i.e. in the earthquake preparation area the current value as well as the direction may change. For example, telluric current may get locked in a vertical plane. It means that in its circuit there may be a segment where the current will be directed from the crack surface, formed in the focus, to the Earth's surface. The possibility of such effect is proved by experimental data (Moroz et al., 2004; Takeuchi et al.,2012).

As a rule, a process of a strong earthquake preparation begins a long earlier then the earthquake and possibly manifests itself in regional foreshocks in a quite long distance from the incoming earthquake focus. Therefore, it seems possible that perturbations of the telluric field might be recorded rather earlier then the incoming earthquake and quite far from its focus (Li, et al.,2013). Thus, perturbation (polarization) of the telluric current, which may be accompanied by generation of electromagnetic waves of ULF diapason, will take place from a certain phase of the earthquake preparation process when increase of the tectonic stress in the rocks causes formation of considerably large cracks  (Kachakhidze et al., 2013).

A significant telluric effect may develop in cases of large foreshocks and aftershocks too. Thus, there is a quite high probability that the perturbation of the telluric current will continue till changings of the tectonic stress take place in the focus (Moroz et al., 2004; Varotsos,2005; Varotsos, et al., 2011; Uyeda, et al., 2000, 2008; Orihara et al., 2012;)

As in the incoming earthquake focus the intensification of electro-telluric effect must be preceded by the formation of thermo-ionized channel with certain direction, it is possible for the ULF emission, generated at this time, to be distributed sectorally on the earth's surface. Such effect, to some extent, indicates to the boundaries of the epicenter area of the incoming earthquake and the direction of the main fault (Varotsos, et al.,1991; Sarlis, N., 1999; Ohta, et al,2001.Varotsos, et al.,, 2006; Li, et al., 2013).

Thus, the process of crack formation and unification into one fault must be accompanied by certain gravitational and electromagnetic phenomena, effects of which may be revealed in the lower atmosphere as well as in the ionosphere. To these phenomena, besides the VLF and ULF emission, we must attribute the existence of IR (infrared) radiation on the top of the atmosphere (Shyh-Chin Lan, et al.,2011; Ouzounov, et al., 2013; Hayakawa, et ail., 2013).

**2.2. Perturbations of the atmospheric electric field.  Cloudiness**

The weather formation and cloud generation process are determined by a joint action of the atmospheric thermodynamic and electric conditions, which is to some extent influenced by process taking place in the Earth. Namely, such influence may be caused by a telluric current in case the electric field connected to it (a field "forced into" the atmosphere) is sufficiently intense. By this time, if it is oriented against the direction of atmospheric electric field corresponding to "fair weather", then local inversion of the atmospheric electric field direction and the oscillation of the atmospheric electric current will be enabled. Clearly revealing the mechanism stimulating this phenomenon may appear especially significant for weather forecasting and management (Chijevsky,2007).

According to the Frenkel theory (Frenkel, 1949; Liperovsky et al., 2011) one of the factor causing the cloud polarization and atmospheric electric field fluctuation is the formation of condensation centers in the clouds oversaturated with water vapour that causes avalanche-like multiplication of the water drops. On the surfaces of these drops some electric charges accumulate, as a reason for generation of which the ionizing effect of space rays in the lower atmosphere is considered (Harrison, 2004;  Harrison et al., 2009),

There are some data stating that in many seismically active regions (the Caucasus, Kamchatka, Portugal) in quite short time intervals, several days or hours before earthquake occurrence, noticeable changes in the atmospheric electric field potential gradient have taken place (Kachakhidze,et al.,2009; Moroz, et ai., 2004; Silva, et al., 2012) Apart from this, before some of the strong earthquakes the local weather changes have been observed, namely, increase in cloudiness and atmospheric discharge intensity (Shou, 1999; . Guo, et al., 2008; Wu et al., 2009; Guangmeng et al.,2013). These papers emphasize connection between atmospheric electric field changings and the meteorological effects, revealing of which is quite a difficult task. Namely, cloud polarization takes place in the gravitational and electric fields, clearly determining of summarized effect of the action of which is a priori impossible on the background of specific atmospheric thermodynamic conditions and highly variable space factors. There is well-known formula for the movement velocity of water drops in vertical plane, which unifies the joint action of gravity and electric forces

$$V = (mg - qE)/6\pi\eta r \qquad (2),$$

where $V$ – is the drop velocity, $m$ – the mass, $q$- is electric charge, $E$– is the electric field stress, $r$ – is the drop radius, $\eta$ – is the kinematic coefficient of the air viscosity, $g$- is the acceleration of the gravity force. The charge of the drop is determined by the formula: $q = \varepsilon_0 \xi r$ , where $\xi$- is the electro-chemical potential of water, $\varepsilon_0$ – the dielectric constant of vacuum.

According to (2) in case the electric force acting on the drop is commensurable to the gravity force, the direction of the drop velocity may become dependent on the value of the electric force. The direction and intensity of the electric field inside the cloud is depended on the quality of the water drop polarization. This, together with thermodynamic conditions is determined by the density and mobility of the charges. Due to permanent fluctuations of the above mentioned electric characteristics the hydrodynamic velocity of the drops must be quite variable. Namely, in the case of equality of the gravity and electric forces by the (2) formula we may determine the minimal value of the characteristic radius of the drop: 0.5 -1 micrometer (Kuznetsov, 2007). A drop of such size, the mass of which is $m = 10^{-14} - 10^{-15}$ gr, will have floated in the cloud. It is obvious that in case the gravity force exceeds the absolute value of the electric force the drop will move vertically downwards.

According to the "fair weather" model when there are neither hail and thunderstorm (cumulus) clouds nor storm and precipitations, considerably light positive ions caused by ionization move to the upper layers of the atmosphere, and heavier negative ions move to the Earth. It is noteworthy that there is another, different, so-called Simpson model, according to which small size drops are charged negatively and large size drops – positively (Kuznetsov, 2007). In both cases, inside the cloud there must be such electric field, direction of which is antiparallel to the orientation of the vertical atmospheric electric field between the Earth's surface and the lower ionosphere.

It is natural that the increase in densities of the atmospheric charges separated from each other assists with polarization of clouds, i.e. strengthens the electric field in the clouds. As this process takes place under conditions of permanent fluctuation it is natural that quasi-stationary thermodynamic conditions and a steady large scale electro-dynamic image cannot be formed in the cloud. However, if the electric field is more or less steady in the cloud, the unsteady but more or less balanced state will probably be established in some span of time. Supposedly, at this time, inversion of the polarization electric field caused by any reason in the space between the upper and lower boundaries of the cloud forms such an unsteadiness, which, together with precipitation, may cause electric discharge inside the cloud as well as between the cloud and the Earth.

Thus, inversion of the atmospheric electric field will contribute to electric charge density variation inside the cloud, which means the generation of atmospheric electric current caused by diffusive or convective processes.

As in the "fair weather" model the sizes of the drops with opposite sign charges are different then their electric mobility will be different as well. Generally, any kind of motion in a cloud, diffusive or convective, will cause generation of atmospheric electric current, which in its turn will influence on the stability of the cloud. Namely, in the case of "fair weather" the intensification of convective motion inside the cloud may become a reason for unsteadiness of the image of distribution of water drops with different mass according to the barometric law.

For more obviousness of this process we may refer to physical analogy between drop movement inside a cloud and convection of water vapour clusters in liquid.

There is a well-known effect of transformation of free convective movement into forced convection, reason of which is the joint action of the gravity field and external heat source (Schlichting,1974). Similar effect may be caused by inversion of the electric field in the cloud that must be stimulated by electric field inversion in a certain segment of the LAI system. Like liquid in stationary free convection mode where heat source has activated, the inverted electric field in the cloud that is in quasi-balanced state will cause the perturbation of the hydrodynamic image of the drop movement. Therefore, together with recombination of charged drops with opposite signs, despite the Coulomb repulsion, confluence of drops with different mass and the same sign may occur in the convection movement process. As a result of the enlargement, due to the reduction of the total area of the drops surface the density of the atmospheric electric charges may increase. Therefore, electric discharge may take place inside the cloud as well as between the cloud and the Earth.

According to one of the modifications of the "fair weather" model, compared to negatively charged drops, positively charged smaller size drops gather in the so-called electro-sphere (Kuznetsov, 2007). In this model the effect of the deceleration of the galactic space rays as the generation source of the atmospheric charge is especially emphasized. It is considered that negatively charged heavy mass drops fall onto the Earth that gains negative polarity in this way. Consequently, as a result of the electric induction effect the lower edge of the cloud will appear to be positively polarized. Therefore, the inversion of the atmospheric electric field, i.e. the variation of the Earth's surface polarity may not occur in ordinary conditions. Thus, in the modified model, like in the Frenkel model, inversion of the electric field may be caused only by variation in the meteorological conditions that depends on thermodynamic properties of the cloud and density of the atmospheric aerosols (Frenkel,1949). Therefore, according to this theory, the velocity changing of the processes of vaporization and condensation inside the cloud and the phenomenon of the electric field inversion appeared to be in the cause-effect relationship.

In this regard a following question arises: is there any alternative to atmospheric electric field inversion mechanism, which would not be immediately connected with the thermodynamic processes inside the cloud? If such mechanism exists, will the atmospheric electric field inversion cause changes in the meteorological condition, i.e. will the cause-effect relationship between these two phenomena become reversible? We suppose that the analogous circuit model (Kachakhidze et al., 2011) gives a quite convincing answer to this question. This model enables to easily determine the physical mechanism that causes the atmospheric electric field inversion effect. Moreover, it is very important that this model is rational from the energy viewpoint. Namely, the earth VLF electromagnetic emission is rarely observed as an isolated electromagnetic impulse and it usually continues for quite a long time.

It is natural that in the same medium conditions it is easier to maintain quasi-stationary Earth electromagnetic emission in a locked system modeled by the analogous circuit. Therefore, we may suppose that the effectiveness of the telluric source of the VLF electromagnetic emission in the atmosphere will increase in case a structure similar to electromagnetic circuit forms in a limited area of space. According to this model the analogous circuit that corresponds to the space between the Earth and the lower ionosphere does not depend on the distance between them. This distance contains the height at which the meteorological condition is formed.

A locked analogous circuit makes the generation of steady electromagnetic waves possible that formally means the existence of favourable conditions from the energy viewpoint. Due to the charged surfaces reflecting the waves the standing electromagnetic waves between the Earth and the ionosphere may exist for quite a long time even in case their generation source is weak or impulsive. Therefore, from the viewpoint of influence on the meteorological condition, we suppose that together with the atmospheric electric field inversion effect, the analysis of the result of influence of the VLF electromagnetic emission is quite interesting.

It is obvious that this issue is a part of the common problem of the natural electromagnetic wave influence on the dissipative dielectric medium of the lower atmosphere. The main source for these waves is the sun and the physical processes stimulated by it. Most of them develop in the main plasmic reservoir of the magnetosphere, plasmosphere and ionosphere – a radiation belt nearest to the Earth. In the formation of the frequency spectrum characteristic of the extremely rare, weakly ionized medium of these structures the various magneto-hydrodynamic and plasmic electromagnetic waves generated during physical processes take place. The frequency spectrum of the latter is in fact continuation of the sun radiation spectrum with the direction to the low frequencies. This frequency diapason begins with magnetospheric VLF electromagnetic radiation, the generation reason for which is considered the cyclotron instability in the electronic component of the magnetospheric plasma medium.

It is known that the sun radiation immediately influences on the magnetospheric processes. In the first place, the result of its activity is ionization of the magnetospheric medium that is the main source for generation of electric charges (Akasofu, Chapman, 1972). Besides ionization, the energy balance in the magnetosphere may change due to the sun radiation and wave-particle type interaction in the magnetospheric medium as well. This effect is supported by the long-wave part of the sun radiation spectrum, the frequency diapason of which covers the frequency spectrum characteristic of the magnetospheric plasma. It is natural that the wave-particle type interaction in the electrically quite neutral lower atmosphere may not develop in the same way as the magnetosphere medium. It is considered that the ionizing action of the galactic space rays will effectively cause changes in the thermodynamic characteristics of the atmosphere only by activation of an alternative source for generation of atmospheric electric charge – radon emanation from the Earth. In such a case the infrared radiation of the sun may have a special influence on the meteorological condition, and consequently, the atmospheric electric field.

There is a theoretical model (Meister et al., 2011, Liperovsky et al., 2008) that connects intensification of the atmospheric infrared radiation with 0.7-20 micrometer wave length and the anomalous changes in the atmospheric electric field value in the epicenter area of the incoming earthquake. This effect inside the cloud is explained on the basis of the Frenkel theory and develops at the background of the electric field direction inversion. This theoretical model requires special conditions, namely, in the epicenter area of the incoming earthquake there must necessarily be sufficient density of the atmospheric aerosols and emanation of radon, as a source of alpha particles, from the Earth. The quasi-stationary activity of this factor must cause anomalously strong pulsations (spikes) of the local atmospheric electric field. Generation of the spikes must continue for a quite long period of time (1-100 minute(s)). The abovementioned paper shows that in the electric field of the spikes with anomalously high tension the atmospheric charged particles may accelerate to such an extent that their energy may reach the necessary limit for generation of infrared radiation.

**2.3. The resonance VLF electromagnetic emission model**

The model (Liperovsky et al., 2008) is quite informative from the viewpoint of presenting the mechanism of infrared radiation generation in the lower atmosphere, though its realization requires satisfaction of quite strict conditions. For example, as it was mentioned above, it is necessary that there was quite intense emanation of radon from the Earth to the atmosphere, and the spike amplitude must reach anomalously high value (1000-3000 V/m) for atmospheric electric field tension. Moreover, the product of the free path of the charged particle ($CO_2$ and $CH_4$) and the electric field intensity and the charge value in the cloud determines the particle energy, which is necessary for generation of infrared radiation with 2-15 micrometer wave length. It is shown that in the case of anomalously strong electric field, for the generation of infrared radiation in a cloud at a normal height, it is sufficient that the minimal length of the free path was ~7 micrometer. It is natural that in case the cloud height reaches 5-10 km the length of the free path will increase due to the decrease of the medium density. Therefore, infrared radiation may be caused by considerably less intensive atmospheric electric field spike. The idea about the variation of the length of the free path according to height in the atmosphere is generally correct.

However, taking into consideration the analogous circuit formalism (Kachakhidze et al., 2011) the value of this parameter may change due to another factor. Here qualitative correction of the (Liperovsky et al. 2008) model is necessary. The issue touches the effect of the atmospheric electric field oscillation in the analogous circuit area, which may be produced in a wide frequency spectrum of the VLF electromagnetic emission: (1 kHz – 1 MHz). Due to different electric mobility of charged particles with opposite signs the fluctuation of the value and direction of the atmospheric electric field support instability of the cloud. This will cause the intensification of the chaotic movement of the charged particles and increase in the length of their free paths. It is expected that this effect will especially increase in the case of considerably high frequency (of MHz order) VLF emission.

According to the analogous circuit model this may occur (Kachakhidze et al., 2011) at a certain stage in the formation of the main fault in the earthquake focus when the characteristic linear scale $l$ of these cracks is minimal yet. Besides, we assume that at such high frequencies in the cloud there may develop a specific resonant effect, which will cause self - generated oscillations of the charged water drops. The basis of this assumption is the well-known theory of hydro-mechanical self- generated oscillation of water drops in the air or, qualitatively the same, air bubbles in water. The reason of these frequencies is the impulsive influence of an external perturbing factor after which the drop elasticity effect caused by the influence of the surface tension force manifests itself. It is known that the main (minimal) frequency of the hydro-mechanical self- generated oscillations of a sphere shape water drop is determined by an expression (Landau, Lifshic,1954)

$$\omega_{min}^2 = 8\alpha/\rho r^3 \qquad (3)$$

where $\alpha$ -is the coefficient of the drop surface tension, $\rho$ -is the water density, $r$ –is the radius of the drop.

The size characteristic of the charged water drops in the cloud changes in a quite big interval: 0.1-50 micrometer, though the value 1-10 micrometer is maximally probable. According to the (3) formula the frequency characteristic of hydro-mechanical self- generated oscillations of such sized clusters must vary in the diapason of ~0.7-25 MHz. In the analogous circuit model the minimum of this diapason is corresponded by $l$=400 m, which in fact is the value characteristic of the crack length (Kachakhidze et al., 2012). On the basis of such analysis we may conclude that inside the cloud formed in the epicenter area of the incoming earthquake a resonance mechanism may activate with the following scheme: in the incoming earthquake focus the necessary conditions for generation of polarization charges are formed. Due to the electric induction between the Earth's surface and the lower ionosphere in this segment of the LAI-system an electromagnetic circuit will be formed and the inversion of the vertical atmospheric electric field will take place. The self-generated oscillations of the circuit are the source for the earth VLF electromagnetic emission. The atmospheric electric field modulated by the frequency of these oscillations will cause hydro-mechanic self-generated oscillations of the charged water drops in the cloud. Consequently, in the cloud the VLF electromagnetic radiation will be generated in MHz frequency diapason. Therefore, additional perturbation (chaotization) of the movement of ions with different electric mobility is expected. This, in its turn, will cause decrease of the free path of the charged particles. It is natural that due to the decrease of the free path length the energy condition for generation of infrared radiation in the cloud will become more severe. This means that at this time the amplitude of the electric field spike may appear insufficient for generation of intensive infrared radiation. It is noteworthy that the considered scheme does not exclude the development of nonlinear resonance effect inside the cloud. Namely, in the case of coincidence of the self-generated oscillation frequencies of analogous circuit and the proper hydro-mechanical self-generated oscillation frequencies of the charged water drops the intensification of the VLF electromagnetic radiation in the atmosphere is possible. Supposedly, the strengthening of this radiation will cause increase of the instability of the cloud and quick confluence of the charged water drops, due to which the meteorological condition will change.

## 2.4. Evaluation of the electromagnetic emission intensity of the telluric current in the geoelectric anomaly area

According to the above mentioned we suppose that the reason of the local inversion of atmospheric electric field changing of the polarity of the Earth's surface segment may be the intensification of the telluric current caused by the activity of the thermoionized channel formed by the earthquake preparing process. The inversion of the atmospheric electric field, according to the analogous circuit model, accompanies the Earth VLF emission. This very phenomenon distinguishes the local segment corresponding to the analogous circuit from the whole LAI system. The latter is formally corresponded by the global electromagnetic circuit, in the self-generated frequency of which, for example, the so-called Schumann resonance takes place.

It is interesting that the ionospheric effect of the VLF emission above seas and oceans is often much stronger than on the land (Hayakawa et al., 2013]. Such circumstances, supposedly, are the result of that the intensity of the telluric current above a sea is usually higher than on the ground. The proof for this is, for example, the paper (Orihara et al., 2012).

In case the focus of the incoming earthquake is under a water layer, supposedly, besides the generation of the ULF diapason waves expressed in geomagnetic field pulsations, the VLF electromagnetic emission must be generated as well. However, in the case of sufficient depth the water layer absorbs the VLF electromagnetic emission and will conduct only ULF diapason waves. But if the telluric current circuit contains a part of the coastal line or shallow, it is possible that the spectrum of the earth electromagnetic emission will be fully revealed.

In any case, there is the probability of electric field image variation in the whole section from the Earth surface to the lower boundary of the ionosphere. Formally, the ionospheric TEC-anomaly may be caused by considerably higher frequency electromagnetic emission generated in the incoming earthquake focus as well as by the geomagnetic pulsations of the ULF emission frequency diapason connected with the telluric current. Namely, activity of this factor may increase the magnetic viscosity of the ionospheric medium. Like the geomagnetic pulsations, the VLF radiation may cause TEC-anomaly due to the local increase of the magnetic viscosity of the plasmic component of the quasi-neutral medium and anomalous "heating up" of the ionosphere (Kereselidze et al., 2009; Hayakawa et al., 2013). Moreover, due to the quite intensive telluric current the change in the Earth's surface polarity may appear a reason for the induction of the TEC-anomaly in the lower ionosphere as well. In the recent years this phenomenon is considered as an especially sensitive indicator for an earthquake preparation process. Therefore, a question about any source of TEC-anomaly genesis arises naturally: does the telluric current have sufficient energy for formation of the local TEC-anomaly in the lower ionosphere? It seems that the answer to this question is quite simple: in the earthquake preparing zone a TEC-anomaly generates, i.e. the Earth VLF electromagnetic emission energy is sufficient for its generation. However, such conclusion is obviously stereotyped and may be acceptable only in case the activity of the space source causing the TEC-anomaly is excluded, though even if such condition is fulfilled the qualitative analysis of the problem will not be sufficient for proving or rejecting the above mentioned idea. Here proper quantitative assessments are required. For this, a quite simple model of the Earth VLF -electromagnetic emission presented below may appear useful.

Let us consider that during the earthquake preparing process an deep generator of polarized charges, a thermo-ionized channel was formed. We may admit that in the polarization area, which we may generally imagine as an electric multipole, an electric current source was activated. Let us call this whole system a geoelectric anomaly. Supposedly, the size characteristic of the telluric current circuit is commensurable with the linear scale of the model multipole. Our goal is evaluation of where the electromagnetic emission energy of the telluric current, a constituting element of a hypothetic geoelectric anomaly, may be sufficient for generation of significant TEC-anomaly in the lower ionosphere. For this purpose it is sufficient to analogize the hypothetic geoelectric anomaly with inhomogeneous space system of polarized charges and use the well-known formulas of classical electrodynamics. The quasi-neutrality of the system requires to fulfill the condition $\tau << T$ where $T$ is the time characteristic of the telluric current variation, $\tau = \frac{l}{c}$ is the time characteristic of the electromagnetic signal distribution, $l$ – the linear size characteristic of the system, $c$ – the light velocity. Generally, the **W** power of the electromagnetic emission of the system is determined by its multipole moments. However, if we require that the magnetic field of the quasi-stationary telluric current obeyed the laws of the stationary field, we should admit that the current circuit is locked in the system area. It seems quite correct in regard to the incoming earthquake focus. In the case of such admission the evaluation of the energy effect of the electromagnetic emission of the system is especially simplified far from the system when the $R' << R$. $R'$-coordinate changes in the limits of the linear scale of the system, and $R$ is the length from the system to the observation point (Stretton, 1948).

In this case, the power of the model system and correspondingly of the electromagnetic emission of the geoelectric anomaly are determined on the lower ionosphere level by the electric and magnetic dipole moments

$$\boldsymbol{P}^e = \int_V \mathbf{R}\rho_0 dV \qquad (4),$$

$$\boldsymbol{P}^m = \frac{1}{2}\int_V [\boldsymbol{R'}, \boldsymbol{J}_0] dV \qquad (5),$$

where $\rho_0$ -is the polarized charge density, $\boldsymbol{J}_0$ – the current.

Thus, the power of the electromagnetic emission of the geoelectric anomaly is determined with quite high precision by the electric dipole moment of the quasi-neutral system of the charges

$$W^e = \frac{\omega^4}{12\pi}\mu\sqrt{\varepsilon\mu}|\boldsymbol{P}^e|^2 \qquad (6),$$

where $\omega$- is the electromagnetic emission frequency, $\varepsilon$ and $\mu$ are the dielectric and magnetic permeability of the medium.

For the specific evaluation of the (6) formula it is necessary to determine the value of the equivalent electric dipole. However, this task is quite difficult as it depends on many parameters: the density of the polarized charge, the electric conductivity of the medium and the sizes of the geoelectric anomaly. But there is some circumstance that enables to simplify the task.

As, according to the model, the telluric current is localized in the geoelectric anomaly area, we may admit that there are conditions necessary for its intensification. At this time the role of the electric current in the radiation loss of the system may significantly increase and the magnetic dipole factor may become commensurable with the electric dipole factor. In such a case the power of the electromagnetic emission of the geoelectric anomaly may evaluated with the following expression

$$W^m = \frac{k^4}{12\pi} \sqrt{\frac{\mu}{\varepsilon}} |P^m|^{(2)} \quad (7)$$

where $k = \frac{2\pi}{l}$ is the wave number, which is determined by the linear size $l$ of the system.

Determining of the magnetic moment of the geoelectric anomaly is as difficult as determining the electric moment. However, if we imagine the equivalent magnetic dipole of the system as the sum of the elementary dipoles then it will be easy to present them in a simple way. For example, we may admit that the telluric current circuit is a unity of separate coils, $r_0$ radius of which is much less compared to the linear size of the geoelectric anomaly. Consequently, the magnetic moment of each coil is $m = \pi r_0^2 I_0$, where $I_0$ - is the telluric current strength. In this case, (7) expression, which corresponds to a single coil, is significantly simplified

$$W^m = 160\pi^6 (r_0/l)^4 I_0^2. \quad (8)$$

By means of the (8) expression for the evaluation of the minimal emission power of the model magnetic dipole let us use the characteristic parameters of the local geoelectric anomaly of Ureki in the coastal line of the Black Sea: the value of the telluric current $I_0 = 0.01$ A (ampere) and the linear scale $l=500$ m (Kereselidze, 2012). Consequently to the variation nature of the specific electric conductivity characteristic of this electric anomaly, we may admit that $r_0=50$ m. In this case the elementary magnetic dipole radiation power corresponding to the telluric current is $16 \times 10^{-4}$ W (watt). It is known that for effective artificial influence on the lower ionosphere medium, minimum 25-100 kW power source is needed (Lichter and all, 1988). Taking this fact into account, it appears that significant perturbation of the ionosphere medium by the influence of the geoelectric anomaly with the given parameters is quite problematic. The above mentioned shows that for formation of the TEC-anomaly the magnetic dipole moment of the telluric current must be, at least, the sum of the elementary magnetic moment corresponding to $10^7$ coils. Consequently, the model presented by us must correspond to a vast geoelectric anomaly, which may form only in a significantly strong earthquake preparation area.

## 3. Conclusion

The Frenkel model of the atmospheric electric field inversion based on the phenomena developed in the external space near the Earth is accepted without alternative so far. We assume that there is one more reason causing the inversion effect, the source of which is the earthquake preparing process.

Activation of the electro telluric factor in the incoming earthquake preparation area may cause formation of a structure similar to the locked electromagnetic circuit in the local sector of the LAI system. Such abstraction (analogous circuit model) simply shows the mechanism of the quasi-stationary Earth VLF electromagnetic emission generation. For its activation the Earth's surface polarity variation is necessary, which means local inversion of the atmospheric electric field direction.

This discussion leads us to the following conclusions:

1. Unlike the Frenkel's model, in the analogous circuit model the reason causing the atmospheric electric field inversion effect is not the phenomena developed in the cloud, but the positive polarization charges accumulated on the Earth's surface; their generation is connected with the telluric phenomena caused by the tectonic processes acting in the incoming earthquake preparation area: the piezo-effect or the electro-kinetic effect.

2. The VLF/LF and ULF electromagnetic emission generated in the earthquake preparation area may become the catalyst for the thermodynamic and electric processes taking place in the lower atmosphere and the reason for the variation of the meteorological conditions caused by instability of the clouds; this effect is mainly revealed during the accumulation of maximal tectonic stress in the incoming earthquake focus, i.e. several days before an earthquake occurrence; in the case of a strong earthquake, the development of these effects during the period of large foreshocks and aftershocks is not excluded.

3. The cloud generation and "bad weather" formation model presented in this work, the cornerstone of which is the atmospheric electric field inversion effect, gives certain theoretical base from the viewpoint of the problem of the local weather management.